\documentclass[preprint,10pt,amsmath,amssymb,floatfix,nofootinbib]{revtex4}                                                

\usepackage{url}
\usepackage{epsf, array, color}
\usepackage{graphicx}
\usepackage{slashed}
\usepackage[linktocpage,colorlinks,urlcolor=blue]{hyperref}

\begin{document}

\newcommand{\vEW}{v_\textrm{EW}}
\newcommand{\thW}{\theta_\textrm{W}}

\title{Constraining New Physics with $h\rightarrow VV$ Tomography}
\author{Matthew Sullivan}
\email{msullivan1@bnl.gov}
\affiliation{Department of Physics, Brookhaven National Laboratory, Upton, New York 11973 U.S.A.}
\date{\today}

\begin{abstract}
The application of quantum information methods to high energy physics has recently been gaining traction. In particular, reconstructing  density matrices and measuring entanglement have been investigated for top quark decays and Higgs decays. This paper will further investigate the utility of density matrices for Higgs decays to vector bosons. Imprints of new physics, whether CP-even or CP-odd, in $h \rightarrow VV$ will generally change the spin density matrix, and so the tomographic reconstruction of the density matrix can constrain, or potentially detect, such new physics. New physics, expressed in the language of the Standard Model effective field theory, is analyzed in this framework of quantum tomography. Prospects for $h \rightarrow WW$ are good due to the fully chiral coupling of the $W$ boson to fermions, while $h \rightarrow ZZ$ requires around an order of magnitude more events to reach comparable sensitivity.
\end{abstract}

\maketitle

\section{Introduction}
\label{sec:intro}

A lot of attention in recent years has focused on the application of quantum information science techniques in collider settings. Recently, entanglement of top quark pairs at the Large Hadron Collider (LHC) by the ATLAS experiment was measured\cite{ATLAS:2023fsd}. Applications of quantum entanglement and quantum tomography to high energy physics are gaining more traction\cite{Afik:2020onf,Fabbrichesi:2021npl,Barr:2021zcp,Severi:2021cnj,Ashby-Pickering:2022umy,Morales:2023gow,Aguilar-Saavedra:2022mpg,Aguilar-Saavedra:2022wam,Aguilar-Saavedra:2024whi,Afik:2022kwm,Barr:2022wyq,Aguilar-Saavedra:2022uye,Altakach:2022ywa,Fabbrichesi:2023jep,LoChiatto:2024dmx,Maltoni:2024tul,Fabbrichesi:2022ovb,Fabbrichesi:2023cev,Dong:2023xiw}.
Understanding the Higgs boson, the only fundamental scalar particle in the Standard Model (SM), and its properties is one of the important goals of the LHC\cite{Dawson:2022zbb}. The Higgs boson's fundamental role in the standard picture of electroweak symmetry breaking leads to predictions of its couplings to other SM particles, particularly to $W$ and $Z$ bosons. Beyond the Standard Model (BSM) physics can change the coupling structure of the Higgs to vector bosons, leading to changes in angular correlations. Measuring deviations from the SM predictions in angular correlations in $h \rightarrow WW$ and $h \rightarrow ZZ$ decays can thus probe BSM physics scenarios.

This paper will outline the application of quantum tomography, the reconstruction of the spin density matrix, to $h \rightarrow WW$ and $h \rightarrow ZZ$ in application to the measurement of heavy BSM physics effects, in terms of Wilson coefficients in the Standard Model effective field theory (SMEFT). Section~\ref{sec:diffdecay} will review how the concept of the the spin density matrix arises for $h \rightarrow VV$ decays. Section~\ref{sec:smeft} will discuss the SMEFT effects in decays of $h \rightarrow VV$. In Section~\ref{sec:tomography}, the techniques for tomographic reconstruction of the spin density matrix from data, and how to constrain the SMEFT Wilson coefficients from it, will be explained.

\section{Differential decay for $h \rightarrow VV \rightarrow 4f$}
\label{sec:diffdecay}

We will consider the decay of $h \rightarrow VV \rightarrow 4f$ at tree level, with both vector bosons $V$ potentially off-shell and neglecting the masses of the fermions $f$. We will additionally assume that the contribution from only one diagram dominates; this can accomplished with phase space cuts or by looking at different flavors of fermions in the final state. First, recall the propagators in the Lorentz gauge for a massive vector boson (including non-zero width),
\begin{equation}
\label{eq:vecprop}
D_{V,\mu \nu}(p) = \frac{-i}{p^2-m_V^2+i m_V \Gamma_V} \left( g_{\mu \nu} - \frac{p_\mu p_\nu}{p^2}\right) ,
\end{equation}
and for the corresponding Goldstone mode,
\begin{equation}
\label{eq:goldstoneprop}
D_G(p) = \frac{-i}{p^2} .
\end{equation}
The numerator in Eq.~\ref{eq:vecprop} projects only the usual three spacelike polarizations (the spin-1 part), whether on-shell or off-shell, while there is generally some remaining timelike polarization (the spin-0 part) when it is off-shell. However, the spin-0 couplings are proportional to the fermion masses, so for massless fermions, only the vector propagator contributes. We can rewrite the numerator of Eq.~\ref{eq:vecprop} in terms of a sum of polarization vectors as
\begin{equation}
\label{eq:vecprop2}
D_{V,\mu \nu}(p) = \frac{-i}{p^2-m_V^2+i m_V \Gamma_V} \sum_\lambda \epsilon^{*}_\mu(p,\lambda) \epsilon_\nu(p,\lambda) .
\end{equation}
Now, we write the decay amplitude in a convenient form:
\begin{widetext}
\begin{eqnarray}
\label{eq:htoVVamplitude}
i \mathcal{M}_{h\rightarrow VV} &=& \sum_{\lambda_1,\lambda_2} i V^{\mu \nu}(p_1,p_2) \epsilon^{*}_\mu(p_1,\lambda_1) \epsilon^{*}_\nu(p_2,\lambda_2)
\frac{-i}{m^{2}_1-m_V^2+i m_V \Gamma_V} \frac{-i}{m^{2}_2-m_V^2+i m_V \Gamma_V} \nonumber \\
&\times&\epsilon_\rho(p_1,\lambda_1) \bar{u}_{s_1}(k_1) (i \gamma^\rho)(\kappa_{1,l} P_L + \kappa_{1,r} P_R) v_{s'_1}(q_1) \nonumber \\
&\times&\epsilon_\sigma(p_2,\lambda_2) \bar{u}_{s_2}(k_2) (i \gamma^\sigma)(\kappa_{2,l} P_L + \kappa_{2,r} P_R) v_{s'_2}(q_2) .
\end{eqnarray}
\end{widetext}
We have written the coupling of the vector particle to fermions in Eq.~\ref{eq:htoVVamplitude} in terms of general left- and right-handed couplings $\kappa_{i,l}$ and $\kappa_{i,r}$, potentially different for the two pairs of fermions, while we have written a general coupling structure $V^{\mu \nu}(p_1,p_2)$ for the $hVV$ coupling. We will focus on looking new physics exclusively on the $hVV$ coupling, so we will always take the $\kappa_{i,l}$ and $\kappa_{i,r}$ couplings as their SM values here.

To see where the spin density matrix and tomography arise, we will proceed to the spin-summed differential decay width,
\begin{widetext}
\begin{eqnarray}
\label{eq:diffwidth}
d\Gamma(h \rightarrow VV) &=& \frac{1}{2 m_h} \sum_{\lambda_1,\lambda_2,\lambda'_1,\lambda'_2}   \frac{dm^{2}_1}{2 \pi} \frac{dm^{2}_2}{2 \pi} \frac{1}{(12 \pi)^2} \frac{(\kappa_{1,l}^2+\kappa_{1,r}^2)m^{2}_1}{(m^{2}_1-m_V^2)^2 - (m_V \Gamma_V)^2} \frac{(\kappa_{2,l}^2+\kappa_{2,r}^2)m^{2}_2}{(m^{2}_2-m_V^2)^2 - (m_V \Gamma_V)^2}\nonumber \\
&\times& d\mathrm{PS}_2(p_H;p_1,p_2) V^{\mu \nu}(p_1,p_2) \epsilon^{*}_\mu(p_1,\lambda_1) \epsilon^{*}_\nu(p_2,\lambda_2) V^{*\alpha \beta}(p_1,p_2) \epsilon_\alpha(p_1,\lambda'_1) \epsilon_\beta(p_2,\lambda'_2) \nonumber \\
&\times& d\mathrm{PS}_2(p_1;k_1,q_1) \mathrm{tr}\left( \slashed{k_1} \gamma^\rho \slashed{q_1} \gamma^\chi (\kappa_{1,l} P_L + \kappa_{1,r} P_R)^2 \right) \epsilon_\rho(p_1,\lambda_1) \epsilon^*_\chi(p_1,\lambda'_1) \frac{12\pi}{(\kappa_{1,l}^2+\kappa_{1,r}^2)m^{2}_1}  \nonumber \\
&\times& d\mathrm{PS}_2(p_2;k_2,q_2) \mathrm{tr}\left( \slashed{k_2} \gamma^\sigma \slashed{q_2} \gamma^\delta (\kappa_{2,l} P_L + \kappa_{2,r} P_R)^2 \right) \epsilon_\sigma(p_2,\lambda_2) \epsilon^*_\delta(p_2,\lambda'_2) \frac{12\pi}{(\kappa_{2,l}^2+\kappa_{2,r}^2)m^{2}_2},
\end{eqnarray}
\end{widetext}
where we have used phase space recursion to write the four-body phase space as a product of integrals over invariant masses, $m_1^2 \equiv p_1^2$ and $m_2^2 \equiv p_2^2$, and two-body phase spaces, $d\mathrm{PS}_2(p_{t};l_1,l_2)$ with total momentum $p_{t}$ and final momenta $l_1$ and $l_2$. Up to normalization, the second line of Eq.~\ref{eq:diffwidth} is the spin density matrix constructed from helicity amplitudes, while the first line contains the proper phase space weighting. The third and fourth lines of Eq.~\ref{eq:diffwidth} are the mappings of the helicity density matrix to final fermion momentum, which will be used for tomography; these lines are normalized so that they integrate to $\delta_{\lambda_i\lambda'_i}$ over the relevant two-body phase space, which is the reason for the appearance of terms with corresponding inverses on the first line. In summary, up to the approximations made, the differential width can be written as the product of a density matrix part, associated with a spin-0 particle decaying to two off-shell spin-1 particles, and a tomographical reconstruction part, associated with the decays of the off-shell spin-1 particles to fermions.

\subsection{Spin density matrix}
Using the first two lines of Eq.~\ref{eq:diffwidth}, we define the non-normalized spin density matrix integrated over phase space as
\begin{widetext}
\begin{eqnarray}
\label{eq:densitymatrix}
\tilde{\rho}_{\lambda_1 \lambda_2}^{\lambda'_1 \lambda'_2} &\equiv& \frac{1}{2 m_h} \int \frac{dm^{2}_1}{2 \pi} \frac{dm^{2}_2}{2 \pi} \frac{1}{(12 \pi)^2} \frac{(\kappa_{1,l}^2+\kappa_{1,r}^2)m^{2}_1}{(m^{2}_1-m_V^2)^2 - (m_V \Gamma_V)^2} \frac{(\kappa_{2,l}^2+\kappa_{2,r}^2)m^{2}_2}{(m^{2}_2-m_V^2)^2 - (m_V \Gamma_V)^2} \nonumber \\
&\times& d\mathrm{PS}_2(p_H;p_1,p_2) V^{\mu \nu}(p_1,p_2) \epsilon^{*}_\mu(p_1,\lambda_1) \epsilon^{*}_\nu(p_2,\lambda_2) V^{*\alpha \beta}(p_1,p_2) \epsilon_\alpha(p_1,\lambda'_1) \epsilon_\beta(p_2,\lambda'_2) ,
\end{eqnarray}
\end{widetext}
where the normalization is such that $\tilde{\rho}_{\lambda_1 \lambda_2}^{\lambda_1 \lambda_2} = \Gamma(h \rightarrow VV \rightarrow 4f)$. We can then define the properly normalized density matrix as 
\begin{equation}
\label{eq:normalizeddensitymatrix}
\rho_{\lambda_1 \lambda_2}^{\lambda'_1 \lambda'_2} \equiv {\tilde{\rho}_{\lambda_1 \lambda_2}^{\lambda'_1 \lambda'_2}}/{\sum_{a,b}(\tilde{\rho}^{ab}_{ab})} .
\end{equation}
The interaction vertex can be written as a sum over Lorentz structures with non-vanishing contributions as
\begin{equation}
\label{eq:vertex}
V^{\mu \nu}(p_1,p_2) = b_1 g^{\mu \nu} + \frac{b_2}{\Lambda^2} \left( p_2^\mu p_1^\nu - p_{1,\alpha} p_2^\alpha g^{\mu \nu}\right) + \frac{b_3}{\Lambda^2} \left( \varepsilon^{\alpha \beta \mu \nu} p_{1,\alpha} p_{2,\beta}\right) ,
\end{equation}
where $\Lambda$ is some mass scale, which will later be taken to be some heavy new physics scale, and $b_i$ are arbitrary coefficients\footnote{One may in general imagine the $b_i$ as functions of the invariant masses $m_1$ and $m_2$ of the vector bosons, but when we match onto SMEFT later, they will just be constants that depend on the Wilson coefficients and other couplings.}. Using this form, we calculate the helicity amplitudes in the Higgs rest frame to be
\begin{widetext}
\begin{equation}
\label{eq:hamplitudes}
\begin{pmatrix}
a_{11} \\
a_{00} \\
a_{-1-1}
\end{pmatrix}
\equiv
\begin{pmatrix}
V^{\mu \nu}(p_1,p_2) \epsilon^{*}_\mu(p_1,1) \epsilon^{*}_\nu(p_2,1)\\
V^{\mu \nu}(p_1,p_2) \epsilon^{*}_\mu(p_1,0) \epsilon^{*}_\nu(p_2,0)\\
V^{\mu \nu}(p_1,p_2) \epsilon^{*}_\mu(p_1,-1) \epsilon^{*}_\nu(p_2,-1)
\end{pmatrix}
=
\begin{pmatrix}
-b_1 +\frac{b_2}{\Lambda^2} \left(m_h^2-m^{2}_1-m^{2}_2\right) + i \frac{b_3}{\Lambda^2} \frac{\lambda^{1/2}(m_h^2,m^{2}_1,m^{2}_2)}{2}\\
b_1 \frac{m_h^2-m^{2}_1-m^{2}_2}{2 m_1 m_2} - \frac{b_2}{\Lambda^2} m_1 m_2\\
-b_1 +\frac{b_2}{\Lambda^2} \left(m_h^2-m^{2}_1-m^{2}_2\right) + i \frac{b_3}{\Lambda^2} \frac{\lambda^{1/2}(m_h^2,m^{2}_1,m^{2}_2)}{2}
\end{pmatrix} ,
\end{equation}
\end{widetext}
where $\lambda^{1/2}(a,b,c)=(a^2+b^2+c^2-2ab-2ac-2bc)^{1/2}$ is the square root of the Källén function. Due to conservation of angular momentum, only these three diagonal helicity amplitudes are non-vanishing. Substituting the helicity amplitudes of Eq.~\ref{eq:hamplitudes} into Eq.~\ref{eq:densitymatrix} and evaluating the trivial two-body phase space integral, we arrive at the following expression for the (non-normalized) spin density matrix in the Higgs rest frame,
\begin{widetext}
\begin{equation}
\label{eq:densitymatrixhelicities}
\tilde{\rho}_{\lambda_1 \lambda_2}^{\lambda'_1 \lambda'_2} = \frac{1}{2 m_h} \int \frac{dm^{2}_1}{2 \pi} \frac{dm^{2}_2}{2 \pi} \frac{\lambda^{1/2}(m_h^2,m^{2}_1,m^{2}_2)}{1152 \pi^3} \frac{(\kappa_{1,l}^2+\kappa_{1,r}^2)m^{2}_1}{(m^{2}_1-m_V^2)^2 - (m_V \Gamma_V)^2} \frac{(\kappa_{2,l}^2+\kappa_{2,r}^2)m^{2}_2}{(m^{2}_2-m_V^2)^2 - (m_V \Gamma_V)^2}  a_{\lambda_1 \lambda_2} a^*_{\lambda'_1 \lambda'_2},
\end{equation}
\end{widetext}
where $a_{\lambda_1 \lambda_2}$ are the helicity amplitudes of Eq.~\ref{eq:hamplitudes}. Note that the density matrix, while ostensibly 9 by 9, only has a 3 by 3 submatrix that is non-zero due to angular momentum conservation.

Rather than just looking at individual matrix entries, it will often be convenient to expand the 3 by 3 submatrices of the density matrix as a sum of Hermitian matrices, which are chosen to be orthonormal under the Hilbert-Schmidt inner product. The set of matrices we will use, which splits into four useful subsets, is
\begin{eqnarray}
\label{eq:matrices}
m^{(1)} &\equiv& \begin{pmatrix}  \frac{1}{\sqrt{3}}  & 0 & 0 \\ 0 & \frac{1}{\sqrt{3}} & 0 \\ 0 &  0 & \frac{1}{\sqrt{3}} \end{pmatrix} , \quad
m^{(2)} \equiv \begin{pmatrix}  \frac{1}{\sqrt{6}}  & 0 & 0 \\ 0 & -2\frac{1}{\sqrt{6}} & 0 \\ 0 &  0 & \frac{1}{\sqrt{6}} \end{pmatrix} , \quad
m^{(3)} \equiv \begin{pmatrix}  0  & \frac{1}{2} & 0 \\ \frac{1}{2} & 0 & \frac{1}{2} \\ 0 &  \frac{1}{2} & 0  \end{pmatrix} , \quad
m^{(4)} \equiv \begin{pmatrix}  0  & 0 & \frac{1}{\sqrt{2}} \\ 0 & 0 & 0 \\ \frac{1}{\sqrt{2}} & 0 & 0  \end{pmatrix} ,
\nonumber \\
m^{(5)} &\equiv& \begin{pmatrix}  \frac{1}{\sqrt{2}} & 0 & 0 \\ 0 & 0 & 0 \\ 0  & 0 & -\frac{1}{\sqrt{2}} \end{pmatrix} , \quad
m^{(6)} \equiv \begin{pmatrix}  0  & \frac{1}{2} & 0 \\ \frac{1}{2} & 0 & -\frac{1}{2} \\ 0 &  -\frac{1}{2} & 0  \end{pmatrix} , \quad
\nonumber \\
m^{(7)} &\equiv& \begin{pmatrix}  0  & \frac{i}{2} & 0 \\ -\frac{i}{2} & 0 & \frac{i}{2} \\ 0 &  -\frac{i}{2} & 0  \end{pmatrix} , \quad
m^{(8)} \equiv \begin{pmatrix}  0  & 0 & \frac{i}{\sqrt{2}} \\ 0 & 0 & 0 \\ -\frac{i}{\sqrt{2}} & 0 & 0  \end{pmatrix} ,
\nonumber \\
m^{(9)} &\equiv& \begin{pmatrix}  0  & \frac{i}{2} & 0 \\ -\frac{i}{2} & 0 & -\frac{i}{2} \\ 0 &  \frac{i}{2} & 0  \end{pmatrix} , \quad
\end{eqnarray}
where the matrices in the first line are real and even under interchange of positive and negative helicities and taking  the transpose,  those in the second line are real and odd under the same transformation, those in the third line are imaginary and even, and the matrix in the last line is imaginary and odd. The real matrices correspond to CP-even physics and the imaginary matrices to CP-odd physics. The matrices that are even under interchanging the helicities and transposing are the only ones that will be necessary for the SMEFT analysis, but the odd matrices are included for completion. Note that $m^{(1)}$ is the only matrix with non-zero trace. 

\subsection{Tomographic information}
We will now review the tomography; a similar discussion can also be found in Ref.~\cite{Aguilar-Saavedra:2024whi}. Our aim is to expand the phase space differential
\begin{widetext}
\begin{eqnarray}
\label{eq:tomography}
dX_{\lambda_1 \lambda_2}^{\lambda'_1 \lambda'_2}&\equiv& d\mathrm{PS}_2(p_1;k_1,q_1) \mathrm{tr}\left( \slashed{k_1} \gamma^\rho \slashed{q_1} \gamma^\chi (\kappa_{1,l} P_L + \kappa_{1,r} P_R)^2 \right) \epsilon_\rho(p_1,\lambda_1) \epsilon^*_\chi(p_1,\lambda'_1) \frac{12\pi}{(\kappa_{1,l}^2+\kappa_{1,r}^2)m^{2}_1} \nonumber \\
&\times& d\mathrm{PS}_2(p_2;k_2,q_2) \mathrm{tr}\left( \slashed{k_2} \gamma^\sigma \slashed{q_2} \gamma^\delta (\kappa_{2,l} P_L + \kappa_{2,r} P_R)^2 \right) \epsilon_\sigma(p_2,\lambda_2) \epsilon^*_\delta(p_2,\lambda'_2) \frac{12\pi}{(\kappa_{2,l}^2+\kappa_{2,r}^2)m^{2}_2},
\end{eqnarray}
\end{widetext}
in two sets of center-of-mass angle variables, where Eq.~\ref{eq:tomography} corresponds to the terms in the third and fourth line of Eq.~\ref{eq:diffwidth}. For the rest frame of the intermediate vector boson $V_1$, we use the polar angle $\theta_1$ and azimuthal angle $\phi_1$ for the fermion momentum $k_1$; for the rest frame of $V_2$, we use $\theta_2$ and $\phi_2$ as measured for the fermion momentum $k_2$. In other words, we will write Eq.~\ref{eq:tomography} as
\begin{widetext}
\begin{equation}
\label{eq:tomographyangles}
dX_{\lambda_1 \lambda_2}^{\lambda'_1 \lambda'_2} = 
F_{\lambda_1 \lambda_2}^{\lambda'_1 \lambda'_2}(\theta_1,\phi_1,\theta_2,\phi_2) d\cos{\theta_1} d\phi_1 d\cos{\theta_2}  d\phi_2
 ,
\end{equation}
\end{widetext}
and find the matrix of functions $F_{\lambda_1 \lambda_2}^{\lambda'_1 \lambda'_2}(\theta_1,\phi_1,\theta_2,\phi_2)$. We will now describe the details of the axes with respect to which these angles are defined.

Let $\hat{z}$ be the unit vector in the direction of the 3-momentum of intermediate vector boson $V_1$ in the Higgs rest frame; note that $-\hat{z}$ is then in the direction of the 3-momentum of $V_2$. Boosts along this axis are used to change between the Higgs rest frame and the $V_1$ or $V_2$ rest frames. Next, we will need to define $\hat{x}$ and $\hat{y}$ in the plane orthogonal to $\hat{z}$; the choice is ultimately arbitrary, but following Ref.~\cite{Aguilar-Saavedra:2024whi}, one can define $\hat{x} = sign(\hat{z} \cdot \hat{p}_p)(\hat{p}_p - \hat{z} (\hat{z} \cdot \hat{p}_p))/ \sqrt{1-(\hat{z} \cdot \hat{p}_p)^2}$ and $\hat{y} = \hat{z} \times \hat{x}$, where $\hat{p}_p$ is the direction of one of the initial state protons as determined in the laboratory frame ($\hat{x}$ and $\hat{y}$ are independent of which proton direction is chosen). $\hat{z}$ will define our polar axis with respect to which $\theta_1$ and $\theta_2$ are measured in the $V_1$ and $V_2$ rest frames, respectively. The azimuthal angles $\phi_1$ and $\phi_2$ are then measured with respect to $\hat{x}$ in the $xy$-plane, again in the $V_1$ and $V_2$ rest frames, respectively.

Because the only non-zero entries in the density matrix were those with matching helicities, we only need to concern ourselves with the entries of Eq.~\ref{eq:tomographyangles} with matching helicities. We will expand Eq.~\ref{eq:tomographyangles} as
\begin{equation}
\label{eq:twobodyexpanded}
F_{\lambda \lambda}^{\lambda' \lambda'}(\theta_1,\phi_1,\theta_2,\phi_2) = \sum_i m_{\lambda \lambda'}^{(i)} f^{(i)}(\theta_1,\phi_1,\theta_2,\phi_2),
\end{equation}
with $\lambda$ and $\lambda'$ not being summed over and with the functions $f^{(i)}(\theta_1,\phi_1,\theta_2,\phi_2)$ defined as
\begin{eqnarray}
\label{eq:PSfunctions}
f^{(1)}(\theta_1,\phi_1,\theta_2,\phi_2) &=& \frac{3 \sqrt{3}}{128 \pi^2} \left( 3 - \cos^2{\theta_1} - \cos^2{\theta_2} - 3 \cos^2{\theta_1}\cos^2{\theta_2} - 4 \zeta_1 \zeta_2 \cos{\theta_1} \cos{\theta_2} \right) \nonumber \\
f^{(2)}(\theta_1,\phi_1,\theta_2,\phi_2) &=& \frac{3 \sqrt{3}}{128 \sqrt{2} \pi^2} \left( 3 - 5 \cos^2{\theta_1} - 5 \cos^2{\theta_2} + 3 \cos^2{\theta_1} \cos^2{\theta_2} + 4 \zeta_1 \zeta_2 \cos{\theta_1} \cos{\theta_2} \right)  \nonumber \\
f^{(3)}(\theta_1,\phi_1,\theta_2,\phi_2) &=& \frac{9}{64 \pi^2} \left( \zeta_1 \zeta_2 - \cos{\theta_1} \cos{\theta_2} \right) \sin{\theta_1} \sin{\theta_2} \cos{\left(\phi_1 - \phi_2 \right)} \nonumber \\
f^{(4)}(\theta_1,\phi_1,\theta_2,\phi_2) &=& \frac{9}{128 \sqrt{2} \pi^2} \sin^2{\theta_1} \sin^2{\theta_2} \cos{2\left(\phi_1 - \phi_2 \right)} \nonumber \\
f^{(5)}(\theta_1,\phi_1,\theta_2,\phi_2) &=& \frac{9}{64 \sqrt{2} \pi^2} \left( -\zeta_1 \cos{\theta_1}\left( 1 + \cos^2{\theta_2} \right) + \zeta_2 \cos{\theta_2}\left( 1 + \cos^2{\theta_1} \right) \right) \nonumber \\
f^{(6)}(\theta_1,\phi_1,\theta_2,\phi_2) &=& \frac{9}{64 \pi^2} \left( \zeta_1 \cos{\theta_2} - \zeta_2 \cos{\theta_1}  \right) \sin{\theta_1} \sin{\theta_2} \cos{\left(\phi_1 - \phi_2 \right)} \nonumber \\
f^{(7)}(\theta_1,\phi_1,\theta_2,\phi_2) &=& \frac{9}{64 \pi^2} \left( \zeta_1 \zeta_2 - \cos{\theta_1} \cos{\theta_2} \right) \sin{\theta_1} \sin{\theta_2} \cos{\left(\phi_1 - \phi_2 \right)} \nonumber \\
f^{(8)}(\theta_1,\phi_1,\theta_2,\phi_2) &=& \frac{9}{128 \sqrt{2} \pi^2} \sin^2{\theta_1} \sin^2{\theta_2} \sin{2 \left(\phi_1 - \phi_2 \right)} \nonumber \\
f^{(9)}(\theta_1,\phi_1,\theta_2,\phi_2) &=& \frac{9}{64 \pi^2} \left( \zeta_1 \cos{\theta_2} - \zeta_2 \cos{\theta_1} \right) \sin{\theta_1} \sin{\theta_2} \sin{\left(\phi_1 - \phi_2 \right)} ,
\end{eqnarray}
where we have defined the shorthand $\zeta_i = (\kappa_{i,l}^2 - \kappa_{i,r}^2)/(\kappa_{i,l}^2 + \kappa_{i,r}^2)$. Because $\phi_1$ and $\phi_2$ only show up through the difference $\phi_1 - \phi_2$, the arbitrary choice of $x$-axis for the definition of the angles earlier does not matter. For the case of $ZZ$ decaying fully leptonically, $\zeta_1=\zeta_2=\frac{1-4\sin^2{\theta_W}}{1-4\sin^2{\theta_W}+8\sin^4{\theta_W}}$, while for all $WW$ decays, $\zeta_1=\zeta_2=1$. However, for semileptonic decays of $WW$ where $\theta_1$ and $\phi_1$ are chosen to be the angles for the hadronically-decaying $W$, unless one can distinguish the jet originating from the quark and the jet originating from the antiquark (such as with charm tagging\cite{Fabbri:2023ncz}), then effectively one has $\zeta_1=0$: any failure to distinguish the particle from the antiparticle is equivalent to losing helicity information about the fermions. For the rest of this paper, we shall assume that the decays are fully leptonic or that the semileptonic $WW$ events being used can be fully reconstructed somehow.

\section{$h \rightarrow VV$ in SMEFT}
\label{sec:smeft}

At this point, we will begin to calculate the density matrices in the context of SMEFT for $h \rightarrow ZZ$ and $h \rightarrow WW$, expanding the properly normalized density matrix to $\mathcal{O}(\Lambda^{-2})$. Of the coefficients $b_i$ in Eq.~\ref{eq:vertex}, only $b_1$ has a non-zero value in the SM; SMEFT corrections to $b_1$ drop out of the properly normalized density matrix at $\mathcal{O}(\Lambda^{-2})$, so we only need to consider the SM value of $b_1$ which we will designate $b_{1,SM}$. We will perform the integration over the entire kinematic range of $m^{2}_1$ and $m^{2}_2$ for demonstration purposes; the density matrix will very close to a pure state and does not change drastically with reasonable phase space cuts.
As discussed earlier, only a 3 by 3 submatrix with matching helicities has non-vanishing entries, so we will show this 3 by 3 submatrix in terms of the matrices in Eq.~\ref{eq:matrices}. The density matrix for $h \rightarrow WW$ is
\begin{eqnarray}
\label{eq:WWdensity}
\rho^{WW} &=& \left(\frac{1}{\sqrt{3}} m^{(1)} - 0.33 m^{(2)} - 0.63 m^{(3)} + 0.28 m^{(4)}\right) \nonumber \\
&+& \left(-1.6 m^{(2)} + 0.69 m^{(3)} - 0.93 m^{(4)}\right)\frac{b^{WW}_2}{b^{WW}_{1,SM}}\frac{(1\textrm{ TeV})^2}{\Lambda^2} \nonumber \\
&+& \left(1.9 m^{(7)} - 1.5 m^{(8)}\right)\frac{b^{WW}_3}{b^{WW}_{1,SM}}\frac{(1\textrm{ TeV})^2}{\Lambda^2},
\end{eqnarray}
while the density matrix for $h \rightarrow ZZ$
\begin{eqnarray}
\label{eq:ZZdensity}
\rho^{ZZ} &=& \left(\frac{1}{\sqrt{3}} m^{(1)} - 0.34 m^{(2)} - 0.63 m^{(3)} + 0.27 m^{(4)}\right) \nonumber \\
&+& \left(-1.4 m^{(2)} + 0.64 m^{(3)} - 0.82 m^{(4)}\right)\frac{b^{WW}_2}{b^{WW}_{1,SM}} \frac{(1\textrm{ TeV})^2}{\Lambda^2} \nonumber \\
&+& \left(1.7 m^{(7)} - 1.3 m^{(8)}\right)\frac{b^{WW}_3}{b^{WW}_{1,SM}} \frac{(1\textrm{ TeV})^2}{\Lambda^2}.
\end{eqnarray}
The exact coefficient in front of $m^{(1)}$ is due to the condition of the density matrix having a trace of 1.

The terms in the SMEFT Lagrangian in the Warsaw basis\cite{Grzadkowski:2010es} that contribute to the density matrices at $\mathcal{O}(\Lambda^{-2})$ in Eq.~\ref{eq:WWdensity} and Eq.~\ref{eq:ZZdensity} are
\begin{widetext}
\begin{eqnarray}
\label{eq:smeftL}
\mathcal{L}_{\mathrm{SMEFT}, \rho} &=& \frac{C_{HW}}{\Lambda^2} H^\dagger H W_{\mu \nu}^I W^{I \mu \nu}
+ \frac{C_{HB}}{\Lambda^2} H^\dagger H B_{\mu \nu} B^{\mu \nu}
+ \frac{C_{HWB}}{\Lambda^2} H^\dagger \tau^I H W_{\mu \nu}^I B^{\mu \nu} \nonumber \\
&+& \frac{C_{H\widetilde{W}}}{\Lambda^2} H^\dagger H \widetilde{W}_{\mu \nu}^I W^{I \mu \nu}
+ \frac{C_{H\widetilde{W}B}}{\Lambda^2} H^\dagger \tau^I H \widetilde{W}_{\mu \nu}^I B^{\mu \nu} 
+ \frac{C_{H\widetilde{B}}}{\Lambda^2} H^\dagger H \widetilde{B}_{\mu \nu} B^{\mu \nu},
\end{eqnarray}
\end{widetext}
where $H$ is the Higgs doublet, $B_{\mu \nu}$ is the $U(1)_Y$ field strength tensor, $W^I_{\mu \nu}$ are the $SU(2)$ field strength tensors, and $\widetilde{B}_{\mu \nu}$ and $\widetilde{W}_{\mu \nu}^I$ are the dual field strength tensors.
It will be convenient to introduce the following shorthand for some combinations of Wilson coefficients that appear:
\begin{widetext}
\begin{eqnarray}
\label{eq:CHZ}
C_{HZ} &\equiv& \cos^2{\thW} C_{HW} +  \cos{\thW} \sin{\thW} C_{HWB} + \sin^2{\thW} C_{HB} \\
\label{eq:CHZtilde}
C_{H\widetilde{Z}} &\equiv& \cos^2{\thW} C_{H\widetilde{W}} +  \cos{\thW} \sin{\thW} C_{H\widetilde{W}B} + \sin^2{\thW} C_{H\widetilde{B}} ,
\end{eqnarray}
\end{widetext}
with $\thW$  the weak mixing angle.
With the operators in Eq.~\ref{eq:smeftL}, we can arrive at the following equations for the coefficients in Eq.~\ref{eq:WWdensity} and Eq.~\ref{eq:ZZdensity},
\begin{eqnarray}
\label{eq:Wparameters}
b^{WW}_{1,SM} &=& \frac{2 m_W^2}{\vEW} \nonumber \\
b^{WW}_2 &=& 4 C_{HW} \vEW \nonumber \\
b^{WW}_3 &=& 4 C_{H\widetilde{W}} \vEW
\end{eqnarray}
\begin{eqnarray}
\label{eq:Zparameters}
b^{ZZ}_{1,SM} &=& \frac{2 m_V^2}{\vEW} \nonumber \\
b^{ZZ}_2 &=& 4 C_{HZ} \vEW \nonumber \\
b^{ZZ}_3 &=& 4 C_{H\widetilde{Z}} \vEW ,
\end{eqnarray}
where $\vEW$ is the Higgs vacuum expectation value, $m_W$ is the mass of the $W$ boson, and $m_Z$ is the mass of the $Z$ boson. With this, we now know the density matrix for $h \rightarrow WW$ and $h \rightarrow ZZ$ to $\mathcal{O}(\Lambda^{-2})$.

\section{Density matrix tomographic reconstruction}
\label{sec:tomography}

Since each entry of the spin density matrix corresponds directly to a particular term in the angular distribution of the fermions, the relationship can be inverted to go from the angular measurements of the events to an estimated density matrix.
We will write the density matrix as the sum
\begin{equation}
\label{eq:DMsum}
\rho_{\lambda \lambda'} = \sum_k \alpha_k m_{\lambda \lambda'}^{(k)} ,
\end{equation}
where $m^{(k)}$ are the hermitian matrices from Eq.~\ref{eq:matrices}. This implies that the angular distribution, according to the tomographic information in Eq.~\ref{eq:twobodyexpanded}, will be 
\begin{equation}
\label{eq:distribution}
\frac{d^{4}\Gamma}{d\cos{\theta_1} d\phi_1 d\cos{\theta_2}  d\phi_2} = \sum_k \alpha_k f^{(k)}(\theta_1,\phi_1,\theta_2,\phi_2) ,
\end{equation}
with the $f^{(k)}(\theta_1,\phi_1,\theta_2,\phi_2)$ as given in Eq.~\ref{eq:PSfunctions}.
Given a set of events with Higgs decaying to two vector bosons, both potentially off-shell, with each vector boson then decaying to fermion pairs, we seek an estimator for the density matrix entries that can be constructed as an average of some function over the set of events. In other words, we seek estimators of the form
\begin{equation}
\label{eq:estimatorform}
\hat{\alpha}_k = \sum_{i=1}^n g^{(k)}(\theta^{(i)}_1,\phi^{(i)}_1,\theta^{(i)}_2,\phi^{(i)}_2),
\end{equation}
where $n$ is the number of events and $\theta^{(i)}_1$, $\phi^{(i)}_1$, $\theta^{(i)}_2$, and $\phi^{(i)}_2$ are the reconstructed angular variables for the $i$th event.

We will present two suitable sets of functions $g^{(k)}(\theta_1,\phi_1,\theta_2,\phi_2)$, one for $WW$ and one for $ZZ$ that can be convolved with a sample of events to estimate the parameters $\alpha_k$ in Eq.~\ref{eq:DMsum}. These particular functions are optimized to minimize the variance of the estimators given the SM density matrix. For more details on the original of these functions, see Appendix~\ref{sec:minvariance}. Suppressing the arguments to all the functions for compactness, for the case of $WW$, we have
\begin{eqnarray}
\label{eq:WWestimators}
g^{WW,(1)} &=& \frac{1}{\sqrt{3}} \nonumber \\
g^{WW,(2)} &=& \left(-0.19 f^{(1)} + 0.95 f^{(2)}+ 0.27 f^{(3)} + 0.030 f^{(4)} \right) / F^{WW,SM} \nonumber \\
g^{WW,(3)} &=& \left(-0.36 f^{(1)} + 0.27 f^{(2)}+ 0.97 f^{(3)} - 1.5 f^{(4)} \right) / F^{WW,SM} \nonumber \\
g^{WW,(4)} &=& \left(0.16 f^{(1)} + 0.030 f^{(2)} - 1.5 f^{(3)} + 9.7 f^{(4)} \right) / F^{WW,SM} \nonumber \\
g^{WW,(5)} &=& \left(0.35 f^{(5)} - 0.41 f^{(6)} \right) / F^{WW,SM} \nonumber \\
g^{WW,(6)} &=& \left(-0.41 f^{(5)} + 1.4 f^{(6)} \right) / F^{WW,SM} \nonumber \\
g^{WW,(7)} &=& \left(0.91 f^{(7)} - 1.5 f^{(8)} \right) / F^{WW,SM} \nonumber \\
g^{WW,(8)} &=& \left(-1.5 f^{(7)} + 9.7 f^{(8)} \right) / F^{WW,SM} \nonumber \\
g^{WW,(9)} &=& 1.7 f^{(9)} / F^{WW,SM}
\end{eqnarray}
where $f^{(i)}$ are the functions from Eq.~\ref{eq:PSfunctions} and $F^{WW,SM}=\left(1/\sqrt{3} f^{(1)} - 0.33 f^{(2)} - 0.63 f^{(3)} + 0.28 f^{(4)} \right)$ is the angular distribution for $h\rightarrow WW$ obtained from the SM density matrix. Similarly for $ZZ$, we have
\begin{eqnarray}
\label{eq:ZZestimators}
g^{ZZ,(1)} &=& \frac{1}{\sqrt{3}} \nonumber \\
g^{ZZ,(2)} &=& \left(-0.20 f^{(1)} + 1.6 f^{(2)} - 0.34 f^{(3)} - 0.26 f^{(4)} \right) / F^{ZZ,SM} \nonumber \\
g^{ZZ,(3)} &=& \left(-0.36 f^{(1)} - 0.34 f^{(2)} + 18 f^{(3)} - 1.1 f^{(4)} \right) / F^{ZZ,SM} \nonumber \\
g^{ZZ,(4)} &=& \left(0.16 f^{(1)} - 0.26 f^{(2)} - 1.1 f^{(3)} + 14 f^{(4)} \right) / F^{ZZ,SM} \nonumber \\
g^{ZZ,(5)} &=& \left(5.4 f^{(5)} - 2.5 f^{(6)} \right) / F^{ZZ,SM} \nonumber \\
g^{ZZ,(6)} &=& \left(-2.4 f^{(5)} + 47 f^{(6)} \right) / F^{ZZ,SM} \nonumber \\
g^{ZZ,(7)} &=& \left(17 f^{(7)} - 1.1 f^{(8)} \right) / F^{ZZ,SM} \nonumber \\
g^{ZZ,(8)} &=& \left(-1.1 f^{(7)} + 14 f^{(8)} \right) / F^{ZZ,SM} \nonumber \\
g^{ZZ,(9)} &=& 44 f^{(9)} / F^{ZZ,SM}
\end{eqnarray}
where now $F^{ZZ,SM}=\left(1/\sqrt{3} f^{(1)} - 0.34 f^{(2)} - 0.63 f^{(3)} + 0.27 f^{(4)} \right)$ is the corresponding SM distribution for $h \rightarrow ZZ$. Note that the structure of which $f^{(i)}$ appear in a given $g^{VV,(j)}$ aligns with the separation of matrices into different lines in Eq.~\ref{eq:matrices}. The main cause of the differences between the case of $WW$ and the case of $ZZ$ is the relative lack of helicity information in $ZZ$. Note that instances of $1/\sqrt{3}$, which come from the density matrix being trace 1 (equivalently, the distribution integrating to 1), are left exact; all other coefficients are calculated numerically.

\section{Extracting Wilson coefficients from the density matrix}
\label{sec:limits}

Since we know from Eqs.~\ref{eq:WWdensity}~and~\ref{eq:Wparameters} how the SMEFT Wilson coefficients affect the $h\rightarrow WW$ density matrix, and similarly from Eqs.~\ref{eq:ZZdensity}~and~\ref{eq:Zparameters} for $h \rightarrow ZZ$, one can extract the Wilson coefficients from the measured density matrices. The Wilson coefficients show up in multiple entries, but one can find the optimal combination for extracting the Wilson coefficient with the minimum variance. This requires knowing the covariance matrix for the density matrix estimators; these are presented in Appendix~\ref{sec:covariance} for the SM, which will be a good approximation to the covariance matrix in SMEFT when the SMEFT contributions are small.

For the case of $WW$, one gets the estimators
\begin{eqnarray}
\hat{C}_{HW} \frac{(1\textrm{ TeV})^2}{\Lambda^2} &=& 8.6 \hat{\alpha}_1^{WW} - 25\hat{\alpha}_2^{WW} + 22\hat{\alpha}_3^{WW} + 2.1 \hat{\alpha}_4^{WW} \nonumber \\
\hat{C}_{H\widetilde{W}} \frac{(1\textrm{ TeV})^2}{\Lambda^2} &=& 30 \hat{\alpha}_7^{WW} + 2.7 \hat{\alpha}_8^{WW} ,
\end{eqnarray}
while for $ZZ$, one similarly gets
\begin{eqnarray}
\hat{C}_{HZ} \frac{(1\textrm{ TeV})^2}{\Lambda^2} &=& -26 \hat{\alpha}_1^{ZZ} - 46\hat{\alpha}_2^{ZZ} + 0.11\hat{\alpha}_3^{ZZ} - 3.4 \hat{\alpha}_4^{ZZ} \nonumber \\
\hat{C}_{H\widetilde{Z}} \frac{(1\textrm{ TeV})^2}{\Lambda^2} &=& 23 \hat{\alpha}_7^{ZZ} - 22 \hat{\alpha}_8^{ZZ},
\end{eqnarray}
where $\hat{\alpha}_i^{VV}$ is the estimator as in Eq.~\ref{eq:estimatorform} using the functions in Eq.~\ref{eq:WWestimators} for $WW$ and the functions in Eq.~\ref{eq:ZZestimators} $ZZ$.
With the covariance matrix for the estimators $\hat{\alpha_i}$, one can also determine the variance of these extracted Wilson coefficients and thus their statistical uncertainties. This leads to being able to answer how many events are needed,  assuming zero background, to reach $\mathcal{O}(1)$ 95\% confidence level error bars for these Wilson coefficients with e.g. a 1 TeV new physics scale. For $C_{HW}$, around 2500 $h \rightarrow WW$ events are needed, and for $C_{H\widetilde{W}}$, around 2400 events are needed\footnote{This is assuming accurate reconstruction of the fermions and antifermions. As touched upon before, for semi-leptonic decays, that means distinguishing quark and antiquark jets. For fully leptonic decays, one also has to accurately reconstruct the momentum of each neutrino.}. The situation for  $h \rightarrow ZZ$ is worse, with around 12000 fully leptonic events being needed for $C_{HZ}$ and 64000 fully leptonic events for $C_{H\widetilde{Z}}$. From the full LHC run 2 data, the number of reconstructed signal events for $h \rightarrow WW \rightarrow l\nu l\nu$ is $\mathcal{O}(1000)$\cite{CMS:2022uhn,ATLAS:2023hyd}, while for $h \rightarrow ZZ \rightarrow llll$, the number of reconstructed signal events is $\mathcal{O}(100)$\cite{CMS:2021ugl,ATLAS:2023tnc}. Considering the effects of backgrounds would lead to even larger statistical uncertainties,  requiring more events to reach the same precision. Even with more than an order of magnitude improvement in statistics over run 2 with the High-Luminosity LHC (HL-LHC), $\mathcal{O}(1)$ sensitivity to TeV-scale new physics in $C_{HZ}$ and $C_{H\widetilde{Z}}$ from Higgs decays appears infeasible at the LHC.

\section{Looking at entanglement measures}
One can look at entanglement in high energy physics for its own sake, but there is also interest in using it to search for new physics. A common method of testing for separability is checking whether the partial transpose (taking the transpose of only one of the systems for a bipartite density matrix) has only non-negative eigenvalues. Although we have used 3 by 3 matrices for simplicity because only a 3 by 3 submatrix of the density matrix has non-zero entries, the full density matrix is in fact 9 by 9, as in Eq.~\ref{eq:normalizeddensitymatrix}.

The partial transpose acts as
\begin{equation}
\label{eq:partialtranspose}
\left(\rho^{PT}\right)_{\lambda_1 \lambda_2}^{\lambda'_1 \lambda'_2} = \rho_{\lambda'_1 \lambda_2}^{\lambda_1 \lambda'_2},
\end{equation}
and given the special form of the density matrix for a spin-0 decay, the nine eigenvalues of $\rho^{PT}$ are easy to find: they are equal to the three diagonal entries and plus or minus the magnitude of the off-diagonal entries for the 3 by 3 submatrix of the density matrix. The off-diagonal entry with the largest magnitude thus serves as a measure of entanglement\footnote{For a completely general system of two qutrits, negativity of the partial transpose is a sufficient, but not a necessary, condition for entanglement. However, for the particular form of the spin-0 density matrix, it is a necessary condition: this measure of entanglement is only zero for diagonal density matrices, and any diagonal density matrix is a convex sum of pure, separable states.}. Expanding consistently to $\mathcal{O}(\Lambda^{-2})$, the magnitudes of the off-diagonal entries for $WW$ using Eq.~\ref{eq:WWdensity} are 
\begin{eqnarray}
\label{eq:entanglementmeasuresWW}
e^{WW}_1 = 0.099 - 0.0041 C_{HW} \frac{(1\textrm{ TeV})^2}{\Lambda^2} \nonumber \\
e^{WW}_2 = 0.039 - 0.0049 C_{HW} \frac{(1\textrm{ TeV})^2}{\Lambda^2} \nonumber \\
e^{WW}_3 = 0.099 - 0.0041 C_{HW} \frac{(1\textrm{ TeV})^2}{\Lambda^2},
\end{eqnarray}
and the corresponding magnitudes of the off-diagonal entries for $ZZ$ using Eq.~\ref{eq:ZZdensity} are
\begin{eqnarray}
\label{eq:entanglementmeasuresZZ}
e^{ZZ}_1 = 0.099 - 0.0032 C_{HZ} \frac{(1\textrm{ TeV})^2}{\Lambda^2} \nonumber \\
e^{ZZ}_2 = 0.039 - 0.0038 C_{HZ} \frac{(1\textrm{ TeV})^2}{\Lambda^2} \nonumber \\
e^{ZZ}_3 = 0.099 - 0.0032 C_{HZ} \frac{(1\textrm{ TeV})^2}{\Lambda^2}.	
\end{eqnarray}

Because the tree-level SM density matrix is CP-even, there are no contributions to the entanglement at $\mathcal{O}(\Lambda^{-2})$ from the CP-odd Wilson coefficients. For $ C_{HW} \frac{(1\textrm{ TeV})^2}{\Lambda^2}$ and $ C_{HZ} \frac{(1\textrm{ TeV})^2}{\Lambda^2}$ of $\mathcal{O}(1)$, the first lines (which are equal to the third lines) of Eq.~\ref{eq:entanglementmeasuresWW} and Eq.~\ref{eq:entanglementmeasuresZZ} dominate, which means the entanglement measure would come from measuring a single density matrix entry. Positive values of $C_{HW}$ ($C_{HZ}$) would then decrease the entanglement in $h \rightarrow WW$ ($h \rightarrow ZZ$), and negative values would increase the entanglement, relative to the SM. Using these entanglement measures to find limits on the CP-even Wilson coefficients, however, leads to worse statistical errors than the measures in Section~\ref{sec:limits} above which are designed to have smaller variances than what one could obtain from any single density matrix entry. We thus reach a similar conclusion to Ref.~\cite{LoChiatto:2024dmx} that entanglement measures don't offer any additional sensitivity to searches for new physics in these particular channels.

\section{Conclusions}
\label{sec:conc}
Spin density matrices offer a new language to discuss new physics contributions in $h \rightarrow VV$. Measuring entanglement in $h \rightarrow VV$ for its own sake would serve as an interesting demonstration of entanglement in a high energy setting, but as an avenue for exploring modifications of the Higgs couplings, the full density matrix holds more promise. Measuring the spin density matrices with quantum tomography corresponds directly to measuring new physics contributions with different Lorentz structures in $hWW$ or $hZZ$ couplings. In terms of SMEFT Wilson coefficients, an $\mathcal{O}(1)$ determination of the CP-even $C_{HW}$  and CP-odd $C_{H\widetilde{W}}$ require around 2400 and 2500 $WW$ events with no background, respectively. The comparable $\mathcal{O}(1)$ determination of the CP-even $C_{HZ}$ and CP-odd $C_{H\widetilde{Z}}$ require around 12000 and 64000 events with no background, respectively. Such a precise measurement for $ZZ$ thus appears beyond even the HL-LHC.

At a future lepton collider, there are more options available. Semi-leptonic and fully hadronic decays are in principle available without the large backgrounds of a hadron collider. Although the coupling of the $Z$ boson to quarks is more chiral, without the ability to distinguish quark jets from antiquark jets, this chirality information is effectively lost. Even without any chiral information, however, it is still possible to measure enough of the spin density matrix to extract the Wilson coefficients. Thus the larger branching ratio of the $Z$ boson to hadrons still makes the hadronic decays useful to improve statistics. Additionally, for $WW$, the semi-leptonic channel also removes the difficulties with reconstructing the momentum of the neutrinos. So even when distinguishing the quark from the antiquark is not possible, $WW$ can also benefit from hadronic decay modes, although the loss of information on the fermion's spin has a greater impact in this case.

\section*{Acknowledgements}
This research is supported by the United States Department of Energy under Grant Contract DE-SC0012704. I would like to thank Haider Abidi, Sally Dawson, Robert Szafron, and Hooman Davoudiasl for useful discussions.

\appendix
\section{Derivation of unbiased estimators with minimum variance for a given distribution}
\label{sec:minvariance}
We will treat the problem as a general problem for some probability distribution which is a linear combination of functions,
\begin{equation}
\label{eq:distsum}
F(X) = \sum_k \lambda_k f_k(X) ,
\end{equation}
where $X$ is any set of kinematic variables and $F(X)$ is a proper probability distribution. We seek an unbiased estimator for each coefficient $\lambda_k$. If we can construct a set of functions $g_l(X)$ such that 
\begin{equation}
\label{eq:dualfunctions}
\int dX f_k(X) g_l(X) = \delta_{kl},
\end{equation}
then we extract $\lambda_l$ using
\begin{equation}
\label{eq:lambdaextraction}
\int dX F(X) g_l(X) = \int dX \sum_k \lambda_k f_k(X) g_l(X) = \lambda_l.
\end{equation}
This would lead directly to an estimator for $\lambda_k$ given by
\begin{equation}
\label{eq:lambdaestimator}
\hat{\lambda}_k = \frac{1}{N} \sum_{i=1}^N g_k(X_i),
\end{equation}
where $N$ is the number of events and the $X_i$ are the kinematic variables corresponding to the $i$th event. Since the kinematic variables are sampled from the distribution of Eq.~\ref{eq:distsum}, the orthogonality relationship of Eq.~\ref{eq:dualfunctions} leads directly to $\langle \hat{\lambda}_k \rangle = \lambda_k$ so that $\hat{\lambda}_k$ is indeed an unbiased estimator.

The problem now is to determine a suitable set of functions $g_l(X)$. There are many such choices of functions, but we can motivate a particular choice. Assume we have a specific hypothesis of distribution which we expect,
\begin{equation}
\label{eq:distsumnull}
F^{(0)}(X) = \sum_k \lambda^{(0)}_k f_k(X) ,
\end{equation}
which we can take as the distribution coming from the SM density matrix. Given the assumption that we are sampling events from this distribution, we can find a set of functions that minimizes the variance as follows.  First, we can write the variance of any of the estimators as
\begin{equation}
\label{eq:estimatorvariance}
\left\langle (\hat{\lambda}_a - \left\langle \hat{\lambda}_a \right\rangle)^2 \right\rangle =
\left\langle \hat{\lambda}_a^2 \right\rangle - \left(\lambda^{(0)}_a\right)^2 =
\int dX \left( F^{(0)}(X) g_a(X)^2\right) - \left(\lambda^{(0)}_a\right)^2.
\end{equation}
Next, we will write a functional for the variance with Lagrange multipliers that enforces the constraint of Eq.~\ref{eq:dualfunctions}:
\begin{equation}
\label{eq:lagrange}
L[g_a(X)] = \int dX \left( F^{(0)}(X) g_a(X)^2\right) - \left(\lambda^{(0)}_a\right)^2 + \xi_{jk} \left(\delta_{jk} - \int dX f_j(X) g_k(X)\right).
\end{equation}
Taking the functional derivative with respect to $g_i(X)$ and setting it equal to zero, we end up with
\begin{equation}
\label{eq:functionalderivative}
\frac{\delta L}{\delta g_i(X)} = 2 F^{(0)}(X) g_i(X) + \xi_{ij} f_j(X) = 0
\end{equation}.
This leads to the conclusion that the estimators that minimize the variance under the assumption of some distribution $F^{(0)}(X)$ are constructed from the functions
\begin{equation}
\label{eq:mvcondition}
g_i(X) = -\frac{1}{2} \xi_{ij} \frac{f_j(X)}{F^{(0)}(X)} ,
\end{equation}
i.e. the functions $g_i(X)$ should be chosen as linear combinations of $f_j(X)/F^{(0)}(X)$ subject to the orthogonality constraint of Eq.~\ref{eq:dualfunctions}.

Note that this choice of functions strictly minimizes the variance only when the distribution really is of the form $F^{(0)}(X)$. However, estimators constructed in this way will always be unbiased estimators of the coefficients $\lambda_a$ regardless of the values of the parameters, and for distributions close to $F^{(0)}(X)$, the variance will still be similar. As an alternative, there is one natural choice of simpler estimators, just by choosing $g_i(X)$ to be linear combinations of $f_j(X)$. For the case of $h \rightarrow WW$, these simpler estimators produced variances about 50\% higher than the estimators designed to minimize the variance given the Standard Model; the difference was milder for $h \rightarrow ZZ$.

\section{Covariance matrices for density matrix estimators}
\label{sec:covariance}
We present here the covariance matrices for the estimators of the coefficients in the density matrix expansion, assuming the SM distribution. As long as the SMEFT corrections are small, the changes to the covariance matrix from SMEFT corrections will also be small. For one event, the covariance matrix for $h\rightarrow WW$ is
\begin{equation}
\label{eq:WWcovarianceSM}
cov\left(\hat{\alpha}_i^{WW}, \hat{\alpha}_j^{WW} \right) = \begin{pmatrix}
 0 & 0 & 0 & 0 & 0 & 0 & 0 & 0 & 0 \\
 0 & 0.84 & 0.063 & 0.12 & 0 & 0 & 0 & 0 & 0 \\
 0 & 0.063 & 0.57 & -1.3 & 0 & 0 & 0 & 0 & 0 \\
 0 & 0.12 & -1.3 & 9.6 & 0 & 0 & 0 & 0 & 0 \\
 0 & 0 & 0 & 0 & 0.35 & -0.41 & 0 & 0 & 0 \\
 0 & 0 & 0 & 0 & -0.41 & 1.4 & 0 & 0 & 0 \\
 0 & 0 & 0 & 0 & 0 & 0 & 0.91 & -1.5 & 0 \\
 0 & 0 & 0 & 0 & 0 & 0 & -1.5 & 9.7 & 0 \\
 0 & 0 & 0 & 0 & 0 & 0 & 0 & 0 & 1.7 
\end{pmatrix},
\end{equation}
while for $h\rightarrow ZZ$ it is
\begin{equation}
\label{eq:ZZcovarianceSM}
cov\left(\hat{\alpha}_i^{ZZ}, \hat{\alpha}_j^{ZZ} \right) = \begin{pmatrix}
 0 & 0 & 0 & 0 & 0 & 0 & 0 & 0 & 0 \\
 0 & 1.5 & -0.55 & -0.16 & 0 & 0 & 0 & 0 & 0 \\
 0 & -0.55 & 18 & -0.90 & 0 & 0 & 0 & 0 & 0 \\
 0 & -0.16 & -0.90 & 14 & 0 & 0 & 0 & 0 & 0 \\
 0 & 0 & 0 & 0 & 5.4 & -2.4 & 0 & 0 & 0 \\
 0 & 0 & 0 & 0 & -2.4 & 47 & 0 & 0 & 0 \\
 0 & 0 & 0 & 0 & 0 & 0 & 17 & -1.1 & 0 \\
 0 & 0 & 0 & 0 & 0 & 0 & -1.1 & 14 & 0 \\
 0 & 0 & 0 & 0 & 0 & 0 & 0 & 0 & 44 
\end{pmatrix} .
\end{equation}
Note that the covariance matrices are block diagonalized according to the separation of the matrices in Eq.~\ref{eq:matrices}; this is part of why expanding in these particular matrices are useful. Note also that the variance of $\hat{\alpha}_1^{WW}$ and $\hat{\alpha}_1^{ZZ}$ is 0; this is due to the density matrices always having trace 1, which is directly enforced by the estimators.

\section{Writing angular variables in terms of Lorentz invariants}
\label{sec:angobservables}
We present here relations between the angles measured in the center-of-mass frame and Lorentz-invariant combinations of the massless fermion 4-momenta here. First, for completeness, since we have the vector boson momenta $p_1 = k_1 + q_1$ and $p_2 = k_2 + q_2$, we have
\begin{eqnarray}
m_1^2 &=& 2 k_1 \cdot q_1 \\
m_2^2 &=& 2 k_2 \cdot q_2 ,
\end{eqnarray}
which can be used to determine the off-shell vector boson invariant masses. Next, we have 
\begin{eqnarray}
(k_1-q_1)\cdot (k_2+q_2) &=& \frac{1}{2} \lambda^{1/2}(m_h^2,m^{2}_1,m^{2}_2) \cos{\theta_1} \\
(k_1+q_1)\cdot (k_2-q_2) &=& -\frac{1}{2} \lambda^{1/2}(m_h^2,m^{2}_1,m^{2}_2) \cos{\theta_2} ,
\end{eqnarray}
which can be used to determine $\cos{\theta_1}$ and $\cos{\theta_2}$. Finally, we have
\begin{eqnarray}
(k_1-q_1)\cdot (k_2-q_2) &=& \frac{1}{2} (m_h^2-m_1^2-m_2^2) \cos{\theta_1} \cos{\theta_2} + m_1 m_2\cos{(\phi_1-\phi_2)} \sin{\theta_1}\sin{\theta_2} \\
\varepsilon_{\mu \nu \rho \sigma} k_1^\mu k_2^\nu q_1^\rho q_2^\sigma &=& \frac{1}{8} m_1 m_2 \lambda^{1/2}(m_h^2,m^{2}_1,m^{2}_2) \sin{(\phi_1-\phi_2)} \sin{\theta_1}\sin{\theta_2} ,
\end{eqnarray}
which can be used to determine the CP-even $\cos{(\phi_1-\phi_2)}$ and CP-odd $\sin{(\phi_1-\phi_2)}$. As explained in the main text, only the difference between the azimuthal angles, $\phi_1-\phi_2$, appears in the angular distribution, and this difference is independent of what axis $\phi_1$ and $\phi_2$ are measured with respect to.

\bibliography{references.bib}

\end{document}